\documentclass[journal]{IEEEtran}

\usepackage{soul}
\soulregister\cite7
\soulregister\ref7
\soulregister\pageref7
\newcommand{\redw}[1]{\textcolor{black}{#1}}

\usepackage{graphicx}
\ifCLASSINFOpdf
\else
\fi

\graphicspath{{figures/}}
\usepackage{tikz}

\usepackage{amsmath,amsfonts,stmaryrd,dsfont}
\usepackage{multirow}

\usepackage[linesnumbered,algoruled]{algorithm2e}

\usepackage{array}

\usepackage{url,cite}

\begin{document}

\title{Model-based STFT Phase Recovery\\ for Audio Source Separation}

\author{Paul~Magron,
	Roland~Badeau,~\IEEEmembership{Senior Member,~IEEE,}
	and~Bertrand~David,~\IEEEmembership{Member,~IEEE}
\thanks{P. Magron is with the Laboratory of Signal Processing, Tampere University of Technology, Finland (e-mail: firstname.lastname@tut.fi), but this work was conducted while he was a PhD student at T\'{e}l\'{e}com ParisTech, Paris, France.

R. Badeau and B. David are with LTCI, T\'{e}l\'{e}com ParisTech, Universit\'{e} Paris-Saclay, 75013, Paris,
France (e-mail: firstname.lastname@telecom-paristech.fr).}
}


\maketitle

\begin{abstract}
For audio source separation applications, it is common to estimate the magnitude of the short-time Fourier transform (STFT) of each source. In order to further synthesizing time-domain signals, it is necessary to recover the phase of the corresponding complex-valued STFT. Most authors in this field choose a Wiener-like filtering approach which boils down to using the phase of the original mixture. In this paper, a different standpoint is adopted. Many music events are partially composed of slowly varying sinusoids and the STFT phase increment \redw{over time} of those frequency components takes a specific form. This allows phase recovery by an unwrapping technique once a short-term frequency estimate has been obtained. Herein, a \redw{novel} iterative source separation procedure is proposed which builds upon these results. It consists in minimizing the mixing error by means of the auxiliary function method. This procedure is initialized by exploiting the unwrapping technique in order to generate estimates that benefit from a temporal continuity property. Experiments conducted on realistic music pieces show that, given accurate magnitude estimates, this procedure outperforms the state-of-the-art consistent Wiener filter.
\end{abstract}

\begin{IEEEkeywords}
Phase recovery, sinusoidal modeling, phase unwrapping, auxiliary function method, audio source separation.
\end{IEEEkeywords}

\IEEEpeerreviewmaketitle

\section{Introduction}

\IEEEPARstart{A}{udio} source separation~\cite{Comon2010} consists in extracting the underlying \textit{sources} that add up to form an observable audio \textit{mixture}. To address this issue, it is common to act on a time-frequency (TF) representation of the data, such as the short-term Fourier transform (STFT), since it provides a meaningful representation of audio signals.

Much research in audio has focused on the processing of nonnegative-valued TF representations, such as the magnitude of the STFT. These representations are usually structured by means of a model, such as nonnegative matrix factorization (NMF)~\cite{Virtanen2007,Fevotte2009}, kernel additive models~\cite{Liutkus2014} or deep neural networks~\cite{Huang2014,Nugraha2016}. However, phase recovery has recently become a growing topic of interest~\cite{Gerkmann2015,Mowlaee2016}. Indeed, obtaining the phase of the corresponding complex-valued STFT is necessary to resynthesize time signals. In the source separation framework, a common practice consists in applying a Wiener-like filtering~\cite{Fevotte2009,Liutkus2015} to the original mixture: the phase of the mixture is then given to each extracted component.
\redw{This technique builds upon the observation that the phase may appear as uniformly-distributed~\cite{Parry2007}, which leads to modeling the complex-valued STFT coefficients as circularly-symmetric random variables (e.g., Gaussian~\cite{Fevotte2009} or stable~\cite{Liutkus2015}). In such a framework, this method yields a set of estimates that is optimal in a minimum mean square error (MMSE) sense.}
However, even if this filter leads to quite satisfactory results in practice~\cite{Virtanen2007,Fevotte2009}, it has been pointed out~\cite{Magron2015} that when sources overlap in the TF domain, it is responsible for residual interference and artifacts in the separated signals.

Improved phase recovery can be achieved with \textit{consistency}-based approaches~\cite{LeRoux2008}. \redw{Consistency is an important property of the STFT since it directly originates from its redundancy property. Indeed, the STFT is usually computed with overlapping analysis windows, which introduces dependencies between adjacent TF bins. Consequently, not all complex-valued matrices are the STFT of an actual time-domain signal. The authors in~\cite{LeRoux2008} proposed an objective function called \textit{inconsistency} that measures this mismatch, and minimizing this criterion results in computing a complex-valued matrix that is as close as possible to the STFT of a time signal.} From the baseline Griffin-Lim (GL) algorithm~\cite{Griffin1984}, several developments have been proposed to design faster procedures~\cite{Zhu2007,Perraudin2013,Beauregard2015,Gnann2010}. For source separation applications, Wiener filtering and consistency-based approaches have been combined in a unified framework~\cite{Gunawan2010,Sturmel2012,Sturmel2013,LeRoux2013}, among which Consistent Wiener filtering~\cite{LeRoux2013} has proved to be the most promising candidate.


Another approach to reconstruct the phase from a spectrogram is to use a phase model based on the observation of fundamental signals that are mixtures of sinusoids~\cite{McAuley1986}. This family of techniques exploits the natural relationship between adjacent TF bins that originates from signal modeling. It leads to a procedure called the~\emph{phase unwrapping} (PU) algorithm, which unwraps the phases over time frames, therefore ensuring the temporal coherence of the signal. \redw{The main difference between the PU algorithm and consistency-based approaches is that the former relies on a signal model while the latter exploit a property of the STFT}. Such an approach has been used in the phase vocoder algorithm~\cite{Laroche1999} for time stretching, and applied to speech enhancement~\cite{Krawczyk2014,Mowlaee2012}, audio restoration~\cite{Magron2015a} and source separation~\cite{Bronson2014}.
\redw{In many cases, the mixtures are assumed to be in harmonic proportions, which means that the partial frequencies are integer multiples of a fundamental frequency, but the PU technique can be extended to signals which do not comply with this assumption~\cite{Magron2015a}.}

In this paper, we introduce a novel source separation procedure which exploits the PU algorithm. Since we focus on the phase recovery issue, the magnitudes are assumed known or estimated beforehand.
We address this problem by considering a cost function which measures the mixing error between the observed and estimated complex mixtures. This function is minimized by means of the auxiliary function method, leading to an iterative procedure. The key idea is to give to the initial estimates the phase obtained with the PU technique. \redw{Indeed, those initial estimates are expected to be close to a local minimum}. Besides, by doing so, the resulting estimates benefit from a temporal continuity property. Experiments conducted on realistic music songs show its potential for an audio source separation task. In particular, it performs similarly to, or better than, the state-of-the-art consistent Wiener filter, with improved interference rejection and a lower computational cost.

This paper is organized as follows. Section~\ref{sec:pu} presents the PU algorithm that is obtained from a sinusoidal model. Section~\ref{sec:ssep} introduces an audio source separation framework which uses this technique. Section~\ref{sec:exp} experimentally validates the potential of the PU algorithm for audio source separation. Finally, section \ref{sec:conclu} draws some concluding remarks.

\section{The phase unwrapping algorithm}
\label{sec:pu}


\subsection{Sinusoidal modeling}
\label{sec:pu_sinus}

Let us consider a mixture of $P$ \redw{time-varying sinusoids:
\begin{equation}
\forall n \in \mathbb{Z} \text{, } x(n) =  \sum_{p=1}^P A_p(n)  e^{2i \pi \nu_p(n) n + i\varphi_{p}},
\label{eq:sinus}
\end{equation}
where $A_p(n)>0$ are the amplitudes, $\nu_p(n) \in [0 , \frac{1}{2}] $ are the normalized frequencies and $\varphi_{p} \in ]-\pi , \pi] $ are the initial phases. Note that we use here complex sinusoids while audio signals are real-valued. However, this is a widely-used model~\cite{McAuley1986,Laroche1999,Lagrange2007} and it yields results similar to the real-valued model since we do not account for negative frequency components.}

\begin{figure}[t]
	\centering
	\includegraphics[scale=0.5]{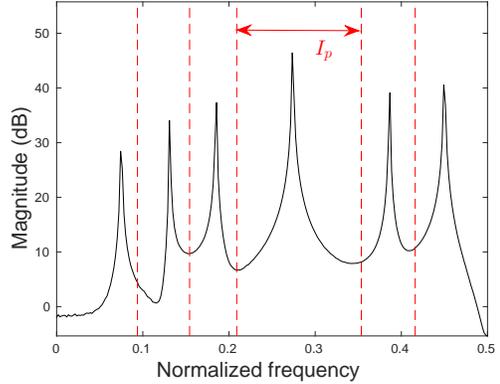}
	\caption{Example of a spectrum (solid line) decomposed into regions of influence (dashed lines).}
	\label{fig:reg_inf}
\end{figure}

\redw{Let $w$ be an analysis window of length $N_w$, which means that $w(n)=0$ $\forall n \notin \{ 0,...,N_w-1 \}$. We assume that the mixture is locally stationary, i.e., that the sinusoidal parameters are constant within time segments of length $N_w$. In order words, for a given time frame $t$, we can note $\nu_p(n+St) = \nu_p(t)$ and $A_p(n+St) = A_p(t)$, where $S$ is the hop size of the STFT. Therefore, the STFT of this mixture in a frequency channel $f$ and time frame $t$ is:
\begin{equation}
X(f,t) = \sum_{p=1}^P A_p(t) e^{2i \pi S \nu_p(t) t + i \varphi_{p}} W_{\nu_p(t)}(f),
\label{eq:stft_mix_sin}
\end{equation}
where $W_{\nu_p(t)}$ is the discrete Fourier transform of the analysis window modulated by the $p$-th frequency in time frame $t$:
\begin{equation}
W_{\nu_p(t)}(f) = \sum_{n=0}^{N_w-1} w(n) e^{2i \pi \left( \nu_p(t)-\frac{f}{N_w} \right) n }.
\end{equation}
}Now, let us assume that there is at most one active sinusoid per frequency channel and per source. Drawing on~\cite{Laroche1999}, we propose to partition the whole frequency range into several regions called \emph{regions of influence}. A region of influence $I_p(t) \subset \{ 0,...,F-1 \}$ corresponds to the set of frequency channels where the STFT is mainly determined by the $p$-th sinusoidal partial, i.e., the contributions of the other partials are negligible:
\begin{equation}
\forall f \in I_p(t) \text{, } X(f,t) = A_p(t) e^{2i \pi S \nu_p(t) t + i \varphi_{p}} W_{\nu_p(t)}(f).
\label{eq:stft_reginf}
\end{equation}
Several definitions of regions of influence exist in the literature, such as choosing their boundaries as the channels of lowest energy between the peaks~\cite{Laroche1999}. \redw{Here we propose to adjust these boundaries so that the greater a magnitude peak is relatively to the neighboring peaks, the wider its region of influence becomes. The resulting regions, whose exact definition can be found in~\cite{Magron2015a}, are illustrated in Fig.~\ref{fig:reg_inf}.}

\redw{From~\eqref{eq:stft_reginf} we obtain the phase of the STFT $\phi = \angle X $, where $\angle$ denotes the complex argument:
\begin{equation}
\forall f \in I_p(t) \text{, } \phi(f,t) = 2 \pi S \nu_p(t) t + \varphi_{p} + \angle W_{\nu_p(t)}(f).
\label{eq:phase_sinus_update_mix}
\end{equation}
Finally, we assume that the sinusoids are \textit{slowly-varying}~\cite{Krawczyk2014}, which means that $\nu_p(t) \approx \nu_p(t-1)$.} This leads to the following recursive relationship between the phase of adjacent time frames:
\begin{equation}
\phi(f,t) \redw{\approx } \phi(f,t-1) + 2 \pi S \nu(f,t).
\label{eq:phase_update_variable_approx}
\end{equation}
where \redw{$\nu(f,t)=\nu_p(t)$ $\forall f \in I_p(t)$ with $ p \in \{ 1,...,P \}$}. This relationship is called \textit{phase unwrapping}.

\subsection{Frequency estimation}

In order to unwrap the phase through~\eqref{eq:phase_update_variable_approx}, one needs to estimate the frequencies $\nu(f,t)$. Many frequency estimation techniques exist, but they generally either require the phase of the STFT (e.g., as in the phase vocoder algorithm~\cite{Laroche1999}) or are restricted to mixtures \redw{whose frequencies are in harmonic proportions} (for instance, sophisticated versions of the harmonic sum or spectral product, such as the PEFAC algorithm~\cite{Gonzalez2014}).

\begin{figure}[t]
	\centering
	\includegraphics[scale=0.5]{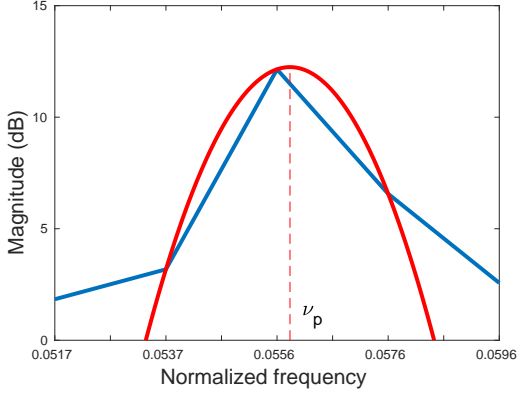}
	\caption{Illustration of the QIFFT technique: a magnitude peak is approximated by a parabola, whose maximum leads to the frequency estimate.}
	\label{fig:qifft}
\end{figure}

Therefore, we have chosen to use the Quadratic Interpolated FFT (QIFFT)~\cite{Abe2004}: we approximate the shape of the log-spectrum around a magnitude peak by a parabola, and the computation of the maximum of the parabola provides a frequency estimate, as illustrated in Fig.~\ref{fig:qifft}. This parabolic approximation is justified theoretically for Gaussian analysis windows, and used in practical applications for any window type. Since this approach provides overall fairly good results~\cite{Betser2008,Abe2004a}, we will use it in our study.

\subsection{The phase unwrapping algorithm}

Algorithm~\ref{al:PU} describes the PU procedure applied in one frame $t$. Note that the algorithm only reconstructs the phase within non-onset frames, since the PU algorithm relies on a recursive relation between adjacent time frames~\eqref{eq:phase_update_variable_approx}. The phase within onset frames must be estimated differently. For instance, in the source separation framework, the mixture phase can be given to each component within onset frames, but alternative estimation techniques (e.g.,~\cite{Magron2015c,Magron2015a}) can be used.

\begin{algorithm}[t]
	\caption{Phase unwrapping}
	\label{al:PU}
			\textbf{Inputs}: Magnitude spectrum $v \in \mathbb{R}_+^{F}$,\\
            phase in the previous time frame $\phi' \in \mathbb{R}^{F}$.\\
			\textbf{Peak localization} $f_p$ from $v$.\\
			\textbf{Frequencies} $\nu_p $ with QIFFT on $v$ around $f_p$.\\
			\textbf{Regions of influence} $I_p$ and $\forall f \in I_p$, $\nu(f)=\nu_p$ .\\
			\textbf{Phase unwrapping} $\phi =\phi' + 2 \pi S \nu$.\\
			\textbf{Output}: $\phi \in \mathbb{R}^{F}$
		
\end{algorithm}

\section{Source separation procedure}
\label{sec:ssep}

In this section, we introduce a source separation procedure that exploits the PU algorithm. More details on the mathematical aspects related to this procedure are presented in a supporting document~\cite{Magron2016a}.

\subsection{Problem setting}

Source separation consists in extracting the $K$ complex components $X_k$ that form a mixture $X$. In this paper, we consider a linear, instantaneous and monaural mixture model: $X = \sum_{k=1}^K X_k$, and \redw{we assume that for all sources $X_k$, $k \in \{1,...,K \}$, a magnitude estimate $V_k$ is available}. Indeed, in this work, we do not tackle the problem of magnitude estimation. Therefore, in our experiences (see Section~\ref{sec:exp}), magnitudes will be assumed known or estimated beforehand on the isolated source spectrograms (as in \textit{informed} source separation~\cite{Liutkus2012a}).

We address this problem by minimizing the mixing reconstruction error, which is given by the following cost function:
\begin{equation}
\mathcal{C}\redw{(\theta)} = \sum_{f,t} |X(f,t)-\sum_k \hat{X}_k(f,t)|^2,
\label{eq:cost_ssep}
\end{equation}
under the constraint $|\hat{X}_k(f,t)|=V_k(f,t)$, and where $\theta = \{\hat{X}_k \text{, } k \in \{ 1,...,K \} \}$. The Wiener filtering estimates:
\begin{equation}
\hat{X_{k}}(f,t) = \dfrac{V_k(f,t)^2}{\displaystyle \redw{\sum_{l=1}^K} V_l(f,t)^2} X(f,t),
\end{equation}
are not a solution to this problem since they do not verify $|\hat{X}_k| = V_k$.
Thus, we propose here to introduce an iterative procedure which provides a novel set of estimates of the sources. Our idea is that a proper initialization of this procedure will provide estimates that benefit from some properties of the initial estimates (\textit{cf}. Section~\ref{sec:ssep_init}).

\subsection{General framework}

In order to minimize~\eqref{eq:cost_ssep}, \redw{we propose to use the auxiliary function method. Indeed, by decorrelating the variables, this technique makes it possible to update them in parallel rather than sequentially (as in the coordinate descent method). This leads to fast procedures and reduces the risk of local minima, so it is well-suited for addressing source separation problems~\cite{Kameoka2009,Gunawan2010,Fevotte2011a}}. Considering a cost function $h(\theta)$, the idea is to introduce a function $g(\theta,\tilde{\theta})$ which depends on some new parameters $\tilde{\theta}$, and verifies:
\begin{equation}
h(\theta) = \min_{\tilde{\theta}}  g(\theta,\tilde{\theta}).
\label{eq:auxiliary_func}
\end{equation}
Such a function is called an~\emph{auxiliary function}. It can be shown (for instance in~\cite{Kameoka2009}) that $h$ is non-increasing under the following update rules:
\begin{equation}
\tilde{\theta} \leftarrow \arg \min_{\tilde{\theta}}  g(\theta,\tilde{\theta}) \text{ and }
\theta \leftarrow  \arg \min_{\theta}   g(\theta,\tilde{\theta}).
\label{eq:up_aux}
\end{equation}

\subsection{Auxiliary function}

Since all TF bins are treated independently, we remove the indexes $(f,t)$ in what follows for more clarity, \redw{which results in seeking to minimize}:
\begin{equation}
h(\theta) = | X - \sum_k \hat{X}_k |^2.
\label{eq:cost_simp}
\end{equation}
To obtain an auxiliary function for~\eqref{eq:cost_simp}, we first introduce the auxiliary variables $\tilde{\theta}= \{ Y_k, k \in k \in \{ 1,...,K \} \} $ such that $\sum_k Y_k = X$. We have:
\begin{equation}
| X - \sum_k \hat{X}_k |^2  = | \sum_k ( Y_k - \hat{X}_k) |^2.
\end{equation}
We then introduce the following nonnegative weights:
\begin{equation}
\lambda_k  = \dfrac{V_k^2}{\sum_l V_l^2},
\label{eq:lambda}
\end{equation}
which leads to:
\begin{equation}
| X - \sum_k \hat{X}_k |^2  = \left| \sum_k \lambda_k \left( \frac{ Y_k - \hat{X}_k}{\lambda_k} \right) \right|^2.
\end{equation}
Since the weights defined in~\eqref{eq:lambda} verify $\sum_k \lambda_k = 1$, we can apply the Jensen inequality to the convex function $z \to z^2$:
\begin{equation}
| X - \sum_k \hat{X}_k |^2  \leq \sum_k  \frac{ | Y_k - \hat{X}_k|^2}{\lambda_k}.
\end{equation}
Thus, $h(\theta) \leq g(\theta,\tilde{\theta})$ with:
\begin{equation}
g(\theta,\tilde{\theta}) = \sum_k  \frac{ | Y_k - \hat{X}_k|^2}{\lambda_k},
\end{equation}
and the problem becomes that of minimizing $g$ under the constraints $\sum_k Y_k = X$ and $\forall k$, $|\hat{X}_k|=V_k$. Let us prove that $g$ is an auxiliary function of the objective cost function \redw{$h$}, i.e., that it satisfies~\eqref{eq:auxiliary_func}. \redw{To do so, we aim to minimize $g$ with respect to $\tilde{\theta}$ under the constraint $\sum_k Y_k = X$, which is equivalent to finding a saddle point for the functional $\mathcal{L}$:
\begin{equation}
\mathcal{L}(\theta,\tilde{\theta},\gamma) = g(\theta,\tilde{\theta}) + \gamma ( \sum_k \bar{Y_k} - \bar{X} ) + \gamma' ( \sum_k Y_k - X ),
\end{equation}
where $\bar{z}$ denotes the complex conjugate of $z$. Note that we need to introduce the constraint by means of the Lagrange multipliers twice: indeed, since the the function $\mathcal{L}$ is not real-valued, we have to treat separately the variables and their complex conjugates. We then calculate the partial derivatives of $\mathcal{L}$ with respect to the complex variables $Y_k$ and $\bar{Y}_k$ (the so-called Wirtinger derivatives~\cite{Bouboulis2010}).
\begin{equation}
\frac{\partial \mathcal{L}}{\partial Y_k}(\theta,\tilde{\theta},\gamma) = \frac{1}{\lambda_k} (\bar{Y}_k-\bar{\hat{X}}_k) + \gamma',
\end{equation}
and
\begin{equation}
\frac{\partial \mathcal{L}}{\partial \bar{Y}_k}(\theta,\tilde{\theta},\gamma) = \frac{1}{\lambda_k} (Y_k-\hat{X}_k) + \gamma.
\end{equation}
Setting those derivative at zero leads to equivalent conditions, therefore we have to solve:}
\begin{equation}
Y_k = \hat{X}_k + \lambda_k \gamma.
\label{eq:up_Y}
\end{equation}
Besides, summing~\eqref{eq:up_Y} over $k$ and using the constraint $\sum_k Y_k = X$ leads to:
\begin{equation}
\sum_k Y_k = \sum_k \hat{X}_k + \gamma \sum_k \lambda_k = X,
\end{equation}
and since the weights $\lambda_k$ add up to $1$, we obtain:
\begin{equation}
\gamma = X - \sum_k \hat{X}_k,
\end{equation}
which leads to:
\begin{equation}
Y_k = \hat{X}_k + \lambda_k (X - \sum_l \hat{X}_l).
\label{eq:up_Y2}
\end{equation}
Thus, $g(\theta,\tilde{\theta})$ is minimized for a set of auxiliary parameters defined by~\eqref{eq:up_Y2}, and it is quite straightforward to observe that it is then equal to $h(\theta)$. This proves that $g$ is an auxiliary function of $h$.

\subsection{Derivation of the updates}

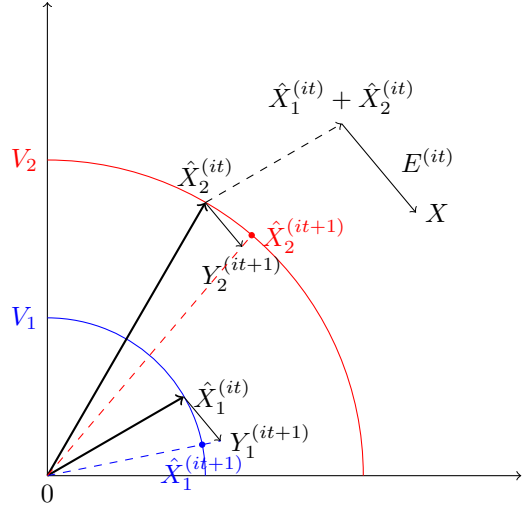
\begin{figure}
	\centering
	\begin{tikzpicture}[scale=0.7]
	\draw[->] (0, 0) node[below]{$0$} -- (9, 0);
	\draw[->] (0, 0) -- (0, 9);
	
	\draw[blue] (3,0)  arc (0: 90: 3) node[left]{$V_1$};
	\draw[red] (6,0)  arc (0: 90: 6) node[left]{$V_2$};
	
	\draw[thick,->] (0, 0) -- (2.59, 1.5) node[right]{$\hat{X}_1^{(it)}$};
	\draw[thick,->] (0, 0) -- (3, 5.19) node[above]{$\hat{X}_2^{(it)}$};
	
	\draw[dashed,->] (3, 5.19) -- (5.59, 6.69) node[above]{$\hat{X}_1^{(it)}+\hat{X}_2^{(it)}$};
	
	\draw[->] (5.59, 6.69) -- (7.0, 5.0) node[right]{$X$};
	\node[right] at (6.54, 6.0) {$E^{(it)}$};
	
	\draw[->] (2.59, 1.5) -- (3.29, 0.66) node[right]{$Y_1^{(it+1)}$};
	\draw[->] (3, 5.19) -- (3.7, 4.35) node[below]{$Y_2^{(it+1)}$};
	
	\draw[dashed, blue] (0, 0) -- (3.29, 0.66);
	\draw[blue, fill=blue] (2.94, 0.59) circle (0.05cm) node[below, very thick]{$\hat{X}_1^{(it+1)}$} ;
	
	\draw[dashed, red] (0, 0) -- (3.88, 4.57);
	\draw[red, fill=red] (3.88, 4.57) circle (0.05cm) node[right, very thick]{$\hat{X}_2^{(it+1)}$} ;

	\end{tikzpicture}
	\caption{Iterative estimation of two complex numbers of fixed magnitude and whose sum is known.}
	\label{fig:estim_phase_mix}
\end{figure}

In accordance with~\eqref{eq:up_aux}, we obtain the update rules on $\theta$ and $\tilde{\theta}$ by alternatively minimizing $g$ with respect to these variables. As it has already been shown, the update rule on $Y_k$ is given by~\eqref{eq:up_Y2}. To obtain the update rule on $\hat{X}_k$, we introduce the constraints $|\hat{X}_k|= V_k$, $\forall k$, by means of the Lagrange multipliers:
\begin{equation}
\mathcal{H}(\theta,\tilde{\theta},\delta_1,...,\delta_K) = g(\theta,\tilde{\theta}) + \sum_k \delta_k (|\hat{X}_k|^2 - V_k^2),
\end{equation}
and we apply the same methodology as before: we compute the zeros of the partial derivatives of $\mathcal{H}$ with respect to the complex variables $\hat{X}_k$. This leads to:
\begin{equation}
\hat{X}_k  =  \frac{Y_k}{1+\lambda_k \delta_k}.
\label{eq:up_X}
\end{equation}
By taking the modulus in~\eqref{eq:up_X} and using the constraints $|X_k| = V_k$, we have:
\begin{equation}
|1+\lambda_k \delta_k| = \frac{|Y_k|}{V_k},
\end{equation}
and finally, by combining this relation and~\eqref{eq:up_X}, we get:
\begin{equation}
\hat{X}_k  = \pm  V_k \frac{Y_k}{|Y_k|}.
\label{eq:up_Xambigu}
\end{equation}
To avoid any ambiguity on the sign in~\eqref{eq:up_Xambigu}, we calculate the value of $g$ for both cases and find that $g$ is minimized with respect to $\theta$ when $\forall k$:
\begin{equation}
\hat{X}_k  =  V_k \frac{Y_k}{|Y_k|}.
\label{eq:up_X2}
\end{equation}
Ultimately, the objective function $h$ is minimized by alternatively applying the update rules~\eqref{eq:up_Y2} and~\eqref{eq:up_X2}. The procedure, illustrated in Fig.~\ref{fig:estim_phase_mix} for $K=2$, actually appears as quite similar to the work in~\cite{Gunawan2010}: the key idea is to distribute the mixing error over the current estimates, and then normalizing. The choice for the weights $\lambda_k$ in~\eqref{eq:lambda} is therefore consistent with this interpretation: the components of highest energy have more impact on the estimation error than the components of lowest energy.

\subsection{Initialization}
\label{sec:ssep_init}

The keystone of our approach is that it enables us to incorporate some prior phase information about the components through a properly-chosen initialization.
Indeed, the cost function $\mathcal{C}$ has many global minima (for $K\geq3$, the problem has infinitely many solutions). Thus, our goal is to find a solution which benefits from some prior knowledge about the phase in order to lead to satisfactorily sounding results.
Intuitively, one could initialize the algorithm by giving the phase of the mixture to each source, but according to~\eqref{eq:up_Y2} and~\eqref{eq:up_X2}, the phase of the estimates would not be modified over iterations.
Then, we propose to initialize this procedure with the PU algorithm: the corresponding initial components are expected to be close to a local minimum and the output estimates can benefit from some temporal continuity. The usefulness of such an initialization will be demonstrated experimentally in Section~\ref{sec:exp_init}.

\subsection{Source separation procedure}

The procedure is summarized in Algorithm~\ref{al:unwrap_mix}. Firstly, let us note that the set of onset frames for each source $\Omega_k$ and the corresponding onset phases $\phi_k^o$ are provided as inputs of the algorithm, since both onset frame detection and onset phase estimation are outside the scope of this article.

Outside onset frames, the phase is initialized by applying the PU technique (Algorithm~\ref{al:PU}). From this initial estimate, we apply the proposed iterative procedure. Finally, we move to the next time frame. A MATLAB implementation of this algorithm is available on the companion website for this paper~\cite{Magron}.

Note that we propose here to apply the iterative procedure within a given time frame before moving to the next one. Alternatively, we can apply the PU algorithm to the whole spectrograms and then apply the iterative procedure in parallel for all TF bins, in order to reduce the computational cost of the method. However, we observed experimentally that the time gain is of about $5 \%$, but the performance in terms of separation quality decreases more significantly. Therefore, it is better to get a good estimate of the phase within a time frame before unwrapping it to the next frame, since it allows us to avoid propagating (and amplifying) the estimation error over time.

\begin{algorithm}[t]
	\caption{Source separation procedure using PU}
	\label{al:unwrap_mix}

			\textbf{Inputs} : \\
			Mixture $X \in \mathbb{C}^{F \times T}$,\\
			Spectrograms $V_k \in \mathbb{R}_+^{F \times T}$, $\forall k \in \{ 1,...,K \}$,\\
			Onset frames sets $\Omega_k$, $\forall k \in \{ 1,...,K \}$, \\
			Onset phases $\phi^o_k(f,t)$, $ \forall t \in \Omega_k$,\\
			Number of iterations $N_{it}$.
			
			\For{$t=1 \text{ to } T-1$}{
			
			\textcolor{gray}{\% Initialization}
             
			\For{$k=1 \text{ to } K$}{
			
			\eIf{$t \in \Omega_k$}
            {Onset phase: $\phi_k(f,t)=\phi^o_k(f,t)$.}
			{$\phi_k(f,t)$ = Phase unwrapping (\textit{cf}. Algorithm~\ref{al:PU}).}

			$\hat{X}_k(f,t) = V_k(f,t) e^{i \phi_k(f,t)}$.
			}
			
            \textcolor{gray}{\% Iterative procedure}
			
			\For{$it=1 \text{ to } N_{it}$}{
			
			 Update $Y_k(f,t)$ with~\eqref{eq:up_Y2},\\
			 Update $\hat{X}_k(f,t)$ with~\eqref{eq:up_X2},

			}
			}
			
			 \textbf{Output} :
			$\forall k \in \{ 1,...,K \}$, $\hat{X}_k \in \mathbb{C}^{F \times T}$.
			
\end{algorithm}

\section{Experimental validation}
\label{sec:exp}


In this section, we evaluate the potential of the PU algorithm for phase recovery and its usefulness for initializing the iterative procedure for source separation. Sound excerpts can be found on the companion website for this paper~\cite{Magron} to illustrate the experiments.

\subsection{Setup}
\label{sec:exp_setup}

We consider $50$ music song excerpts from the DSD100 database, a semi-professionally mixed set of music song used for the SiSEC 2016 campaign~\cite{Liutkus2017}. Each excerpt is $10$ seconds-long and is made up of $K=4$ sources: \texttt{bass}, \texttt{drums}, \texttt{vocals} and \texttt{other} (which may contain various instruments such as guitar, piano...). The signals are sampled at $F_s=44100$ Hz and the STFT is computed with a Hann window, $75$~\% overlap and no zero-padding. The length of the analysis window will be discussed in Section~\ref{sec:gl_vs_pu}.

The \textsc{MATLAB} Tempogram Toolbox~\cite{Grosche2011} provides a fast and reliable onset frames detection from spectrograms (it estimates the onsets before several post-processing operations to find the tempo). The phases within onset frames will be either assumed known or estimated by assigning the mixture onset phase to each source (in Section~\ref{sec:exp_other}). In Section~\ref{sec:exp_onset}, we will investigate the impact of the onset phase estimation on the separation quality.

Finally, the magnitude peaks $f_p$ are tracked from the spectra by using the corresponding MATLAB function (\texttt{findpeaks}) in Algorithm~\ref{al:PU}, and the iterative procedure in Algorithm~\ref{al:unwrap_mix} uses $50$ iterations.

In order to measure the performance of the methods, we use the \textsc{BSS Eval} toolbox~\cite{Vincent2006} which computes various energy ratios: the signal-to-distortion, signal-to-interference, and signal-to-artifact ratios (SDR, SIR and SAR), which are expressed in dB and where only a rescaling (not a refiltering) of the reference is allowed~\cite{Isik2016}.

\subsection{Separation scenarios}
\label{sec:exp_scenarios}

In this paper, we do not address the task of blind magnitude estimation, \redw{therefore the magnitudes are either assumed known or estimated beforehand}.

First, in the oracle scenario, we assume that the magnitudes $V_k$ are equal to the ground truth.

Second, we consider an informed scenario, which corresponds to a \textit{coding-based informed source separation} framework~\cite{Liutkus2012a}: in this scenario, some side-information can be computed from the isolated sources (the \textit{encoding} stage) and then used to enhance the separation performance (the \textit{decoding} stage). A common approach consists of computing a nonnegative matrix~\cite{Rohlfing2016,Rohlfing2016a} or tensor~\cite{Rohlfing2017,Liutkus2013,Ozerov2013} factorization on the isolated source spectrograms and then using the corresponding decomposition to estimate a Wiener filter at the decoding stage. Those approaches have shown very good results in terms of separation quality at a very low bitrate. We therefore propose to apply an NMF with Kullback-Leibler divergence~\cite{Virtanen2007} to the spectrogram of each isolated source, in order to obtain an estimate of the magnitudes $V_k$. Each NMF uses $200$ iterations of multiplicative update rules and a rank of factorization set at $50$. This scenario informs us about the potential of the proposed method for an audio source separation task when the magnitude estimates differ from the ground truth, while still remaining of relatively good quality.

\redw{Third, we consider a blind scenario, in which the magnitude spectrograms are directly estimated from the mixture. We apply $200$ iterations of multiplicative updates of Kullback-Leibler NMF with a rank of factorization set at $200$ to the mixture's spectrogram $|X|$. Then, the NMF components are grouped into $4$ sources by means of the source filter-based clustering method described in~\cite{Spiertz2009} which yields an estimate of the magnitudes $V_k$. This scenario informs us about the impact of phase recovery on source separation quality when the magnitudes are not accurately estimated.}


\subsection{Griffin-Lim vs. phase unwrapping}
\label{sec:gl_vs_pu}

The goal of this experiment is to compare the performance of a consistency-based approach (the GL algorithm) and a model-based approach (the PU algorithm) for a blind phase retrieval task. \redw{Since the songs from the DSD100 database are made up of sources that overlap in the TF domain, we can no longer assume that there is at most one frequency component per channel in the mixtures, which is a key hypothesis in the PU technique (see Section~\ref{sec:pu_sinus}). Therefore, we report here the results for $30$ piano pieces from the MAPS~\cite{Emiya2010a} database, where this scenario is less likely to occur (similar results have been obtained on guitar and speech signals, which we omit here for brevity). Note that for source separation applications, since the PU technique is performed on isolated sources, this will no longer be a problem.}

\begin{figure}[t]
	\hspace{-1em}
	\includegraphics[scale=0.48]{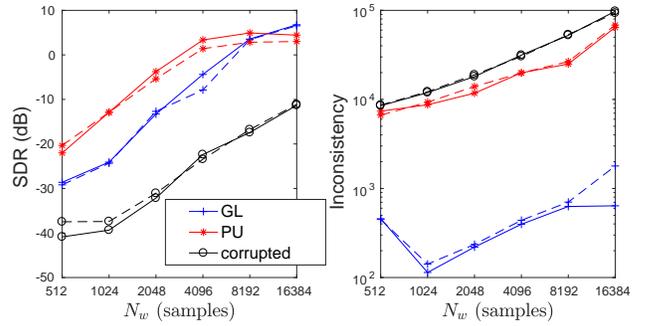}
	\caption{Comparison between GL and PU in terms of \redw{SDR (left) and inconsistency (right) in the oracle (solid lines) and informed (dashed lines) scenarios} for various window lengths.}
	\label{fig:gl_vs_pu}
\end{figure}

First, a comparison between three analysis windows (Hann, Hamming and Blackman) showed no significant difference in terms of SDR. In addition, overlap ratios higher than $75$ $\%$ did not improve the results, while \redw{they required more processing time}. For those reasons, we chose a Hann window with $75$ $\%$ overlap in our experiments. We study here the impact of the window length on the reconstruction quality measured by means of the SDR, and we also compute the inconsistency of the estimates defined as follows:
\begin{equation}
\mathcal{I}(X) = \sum_{f,t} \left| X(f,t)-\text{STFT} \circ \text{iSTFT} (X)(f,t) \right |^2.
\label{eq:inconsistency}
\end{equation}
In this experiment, the onset phases are assumed known. We corrupt the complex STFT of the signals by setting the phases within non-onset frames to random values taken in $]-\pi;\pi]$. We then apply the algorithms in both oracle and informed scenarios. The GL algorithm uses $200$ iterations (performance is not further improved beyond). 
We also report the scores (SDR and inconsistency) computed on the corrupted STFTs as a comparison reference. The results are presented in Fig.~\ref{fig:gl_vs_pu}.

We observe that overall, the PU algorithm outperforms the traditional GL method in terms of SDR for most window lengths. Both algorithms are sensitive to the accuracy of the magnitude spectrogram, as suggested by the drop in SDR values when going from the oracle to the informed scenario for most analysis windows. However, when the spectrogram is no longer equal to the ground truth, our approach still provides overall better results than the consistency-based GL algorithm.

Both algorithms decrease the inconsistency compared to the corrupted reference, but the GL algorithm performs better than PU according to this criterion. This was expected since the GL algorithm is designed to directly minimize this criterion. However, a comparison between SDR and inconsistency shows that minimizing the inconsistency does not imply increasing the SDR. This suggests that the direct optimization of the inconsistency criterion may not be the most appropriate way of accounting for this property. Besides, the GL algorithm computes an estimate that is optimal (in terms of inconsistency) only locally, while a global minimum would correspond to a null inconsistency in the oracle scenario.

We note that the longer the window, the better the results. Actually, for very long analysis windows, GL performs better than PU in terms of SDR. This may be explained by some artifacts that appear in the signals estimated with the PU technique. Indeed, perceptually (sound examples are available in~\cite{Magron}), two phenomena characterize them: \textit{musical noise}, which appears for short windows, when the frequency resolution is poor, and \textit{phasiness}~\cite{Laroche1999}, which appears for long windows, when the temporal resolution is poor\footnote{To overcome the issue of looking for a compromise between temporal and frequency resolution, a multiple resolution framework could further be investigated, as in some improved versions of the phase vocoder~\cite{Roebel2003a,Juillerat2017}.}. \redw{In the latter case, the local stationarity assumption, on which the PU algorithm is based, does no longer hold, thus leading to a decrease of its performance.} Therefore, it is not obvious that the SDR is able to capture both the musical noise and the phasiness phenomena. Indeed, some informal listening tests showed that windows shorter than $16384$ samples lead to more satisfactorily sounding results for both algorithms, while they make the SDR decrease according to Fig.~\ref{fig:gl_vs_pu}. In particular, a $4096$ sample-long analysis window leads to the best results in terms of perceptual quality, with a fairly high SDR.

Finally, for audio source separation applications, we tested different window lengths and we observed that a $4096$ sample-long analysis window leads to the best results in terms of SDR for the consistent Wiener filtering technique and the proposed iterative procedure. Therefore, we use this value in the following experiments.

\subsection{Initialization of the iterative procedure}
\label{sec:exp_init}

Here, we investigate the influence of the initialization of the iterative procedure on the separation quality, as motivated in Section~\ref{sec:ssep_init}. Let us consider $50$ songs from the DSD100 dataset in the oracle scenario, and the onset phases are assumed known. We run the procedure with different initializations (line 13 in Algorithm~\ref{al:unwrap_mix}): either PU or random values. \redw{We also test initializing with the mixture phase in order to obtain a comparison reference, even if the phase of the estimates will not be modified over iterations, as explained in Section~\ref{sec:ssep_init}.}
The results provided in Table~\ref{tab:source_sep_init} show that the initialization with the PU algorithm significantly improves the separation quality over the other initializations.

\begin{table}[t]
	\center
		\caption{Source separation performance (SDR, SIR and SAR in dB) for various initializations on the DSD100 dataset.}
		\label{tab:source_sep_init}
	\begin{tabular}{c|ccc}
		Initialization & SDR & SIR & SAR \\
		\hline
        Mixture & $7.5$ & $13.7$ & $8.9$\\
		Random  & $9.5$ & $22.8$ & $9.7$\\
		PU & $\mathbf{13.6}$ & $\mathbf{31.0}$ & $\mathbf{13.7}$ \\
		\hline
	\end{tabular}
\end{table}

\begin{figure}[t]
	\centering
	\includegraphics[scale=0.5]{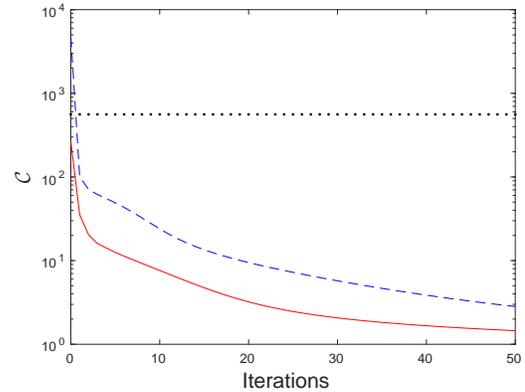}
	\caption{Error \redw{$\mathcal{C}$} over iterations within a TF bin where the sources overlap. \redw{The dotted and solid lines respectively correspond to the initializations with the mixture phase and PU algorithm}, and the dashed line corresponds to the average values over $10$ random initializations.}
	\label{fig:ssep_init_dsd_error}
\end{figure}

We consider one mixture and we plot the error \redw{$\mathcal{C}$} in a TF bin where the sources overlap in Fig.~\ref{fig:ssep_init_dsd_error}. We see that the PU initialization leads to a better and faster convergence (in terms of error) than a random initialization. \redw{Besides, these two techniques reach a significantly lower error value than the initialization with the mixture phase. Perceptually (audio examples are available at~\cite{Magron}), we observe that the PU initialization yields estimates with less artifacts than the other techniques, especially in the \texttt{bass} and \texttt{drum} tracks (which overlap the most). We also note that there is no significant difference in terms of sounding quality between a random initialization and using the mixture's phase, while the corresponding output errors (see Fig.~\ref{tab:source_sep_init}) strongly differ. This suggests that this criterion does not perfectly retrieve perceptual criteria.}

\begin{figure}[t]
	\centering
	\includegraphics[scale=0.5]{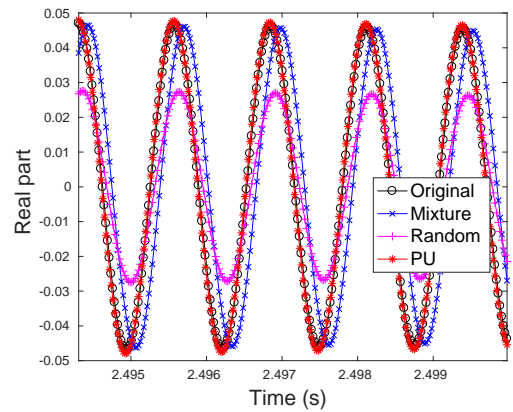}
	\caption{Real part of the third partial ($784$ Hz) \redw{in the STFT channel at $786$ Hz} of a C4 piano note where it overlaps with another note (G4), for various initializations of Algorithm~\ref{al:unwrap_mix}.}
	\label{fig:ssep_init_piano}
\end{figure}

To illustrate this result, we consider a mixture composed of two piano notes from the MAPS database. In order to visualize the real parts of the reconstructed components, we synthesize time-domain signals and compute another STFT with a hop size of $1$ sample. As illustrated in Fig.~\ref{fig:ssep_init_piano}, the initialization with PU yields components that better fit the original signal compared to the other approaches. This confirms the usefulness of the PU algorithm to initialize Algorithm~\ref{al:unwrap_mix}.

\subsection{Comparison to other methods}
\label{sec:exp_other}

\begin{figure}
	\hspace{-2em}
	\includegraphics[scale=0.8]{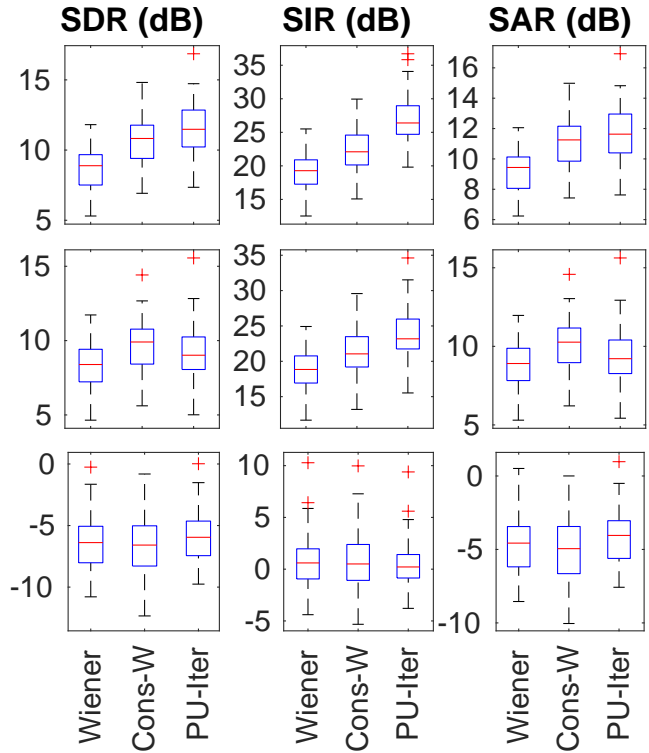}
    \vspace{-1cm}
	\caption{Source separation performance of various methods in the oracle (top), informed (middle) and blind (bottom) scenarios. Each box-plot is made up of a central line indicating the median of the data, upper and lower box edges indicating the $1^{st}$ and $3^{rd}$ quartiles, whiskers indicating the minimum and maximum values, and crosses representing the outliers.}
	\label{fig:ssep_bss}
\end{figure}

In this experiment, the onset phases are estimated by giving the mixture phase to each component. We compare the following methods: Wiener filtering~\cite{Fevotte2009}, consistent Wiener filtering~\cite{LeRoux2013}\footnote{This technique depends on a weight parameter that promotes the consistency constraint. \redw{It is learned beforehand on $50$ other songs from the dataset by choosing the value that maximizes the SDR, SIR and SAR.}}, and Algorithm~\ref{al:unwrap_mix}. Those methods will be respectively denoted \textbf{Wiener}, \textbf{Cons-W} and \textbf{PU-Iter}. The separation is performed on the $50$ songs composing the dataset, and the results are represented with box-plots in Fig.~\ref{fig:ssep_bss}. To complete these results, we also provide some sound excerpts~\cite{Magron} so that the interested reader can assess the sounding quality of the corresponding estimates.

In the oracle scenario, \textbf{PU-Iter} outperforms \textbf{Wiener} and \textbf{Cons-W}, notably in terms of SIR. In addition to those indicators, we perceptually note that \textbf{Cons-W} and \textbf{PU-Iter} lead to similar \texttt{vocals} and \texttt{other} tracks (in terms of sounding quality), but the \texttt{bass} and \texttt{drum} tracks estimated with \textbf{PU-Iter} are of higher quality: the \texttt{bass} is neater and the musical noise artifacts in the \texttt{drum} track are reduced compared with the other approaches.

In the informed scenario, the proposed method yields slightly worse results than \textbf{Cons-W} in terms of SDR and SAR (it is still better than \textbf{Wiener}), but leads to an improvement in terms of interference rejection. This observation is consistent with previous works~\cite{Magron2016} on sinusoidal model-based phase recovery, where this approach has been shown useful to reduce the interference at the cost of more artifacts. However, when listening at the corresponding excerpts, we observe that the \textbf{PU-Iter} method still enhances the sounding quality of the \texttt{bass} track compared to the \textbf{Cons-W} technique, while other tracks are similar.

\redw{In the blind scenario, all methods' performance significantly decrease compared to the other scenarios, and yield overall similar results. Nonetheless, \textbf{PU-Iter} leads to a slight increase in SDR and SAR compared to the other methods, but this method's interest is greater when the magnitude spectrograms are reliably estimated.}

\begin{table}[t]
	\center
    	\caption{Average PEASS scores.}
        \label{tab:sep_peass}
	\begin{tabular}{ll|ccc}
		 & & Wiener & Cons-W & PU-Iter \\
		\hline
        \multirow{4}{*}{Oracle}
		&  OPS & $19.2$  & $19.7$ & $\textbf{27.4}$  \\
		&  TPS & $28.4$  & $30.4$ & $\textbf{36.1}$  \\
		&  IPS & $34.7$  & $34.5$ & $\textbf{39.8}$  \\
		&  APS & $30.6$  & $31.0$ & $\textbf{36.3}$   \\
		\hline
        \multirow{4}{*}{Informed}
		&  OPS & $19.0$  & $19.0$ & $\textbf{24.5}$  \\
		&  TPS & $25.8$  & $26.9$ & $\textbf{33.3}$  \\
		&  IPS & $34.0$  & $33.8$ & $\textbf{41.5}$  \\
		&  APS & $28.8$  & $28.9$ & $\textbf{33.5}$   \\
        \hline
        \multirow{4}{*}{Blind}
		&  OPS & $10.4$  & $10.5$ & $\textbf{12.5}$  \\
		&  TPS & $11.6$  & $12.4$ & $\textbf{17.0}$  \\
		&  IPS & $33.1$  & $\textbf{32.6}$ & $27.7$  \\
		&  APS & $17.5$  & $18.2$ & $\textbf{25.4}$   \\
		\hline
	\end{tabular}
\end{table}

Since our informal listening evaluation is somewhat inconsistent with the results in terms of objective criteria, we propose to compute the PEASS score~\cite{Emiya2011}, which provides a novel set of criteria that is built upon a subjective evaluation of source separation quality, and designed to better match perception than the SDR, SIR and SAR. The resulting criteria are the overall, target-related, interference-related and artifacts-related Perceptual Scores (OPS, TPS, IPS and APS). The corresponding results are presented in Table~\ref{tab:sep_peass}. We observe that for all those criteria, in both the oracle and informed scenario, the proposed \textbf{PU-Iter} method outperforms \textbf{Wiener} and \textbf{Cons-W} by a large margin: in particular, the improvement when going from \textbf{Cons-W} to \textbf{PU-Iter} is more significant than the improvement of \textbf{Wiener} over \textbf{Cons-W} on average. This confirms the performance of the proposed approach and is consistent with our informal perceptive evaluation. \redw{In the blind scenario, \textbf{PU-Iter} still outperforms the other techniques, except in terms of IPS. Overall, the relative performance of \textbf{PU-Iter} increases when the magnitude estimates get close to the ground truth. Indeed, the OPS difference between \textbf{PU-Iter} and \textbf{Cons-W} is of $2$, $5.5$ and $7.7$ in the blind, informed and oracle scenarios respectively. 
This confirms that while the proposed approach shows good results compared to other methods, its potential is fully exploited when the magnitude spectrograms are accurately estimated.}

\begin{figure}[t]
	\centering
	\includegraphics[scale=0.5]{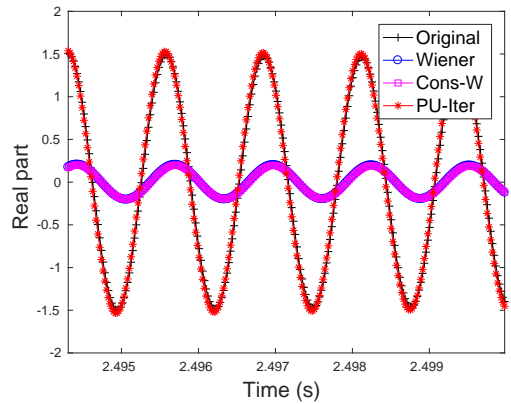}
	\caption{Real part of the third partial ($784$ Hz) \redw{in the STFT channel at $786$ Hz} of a C4 piano note where it overlaps with another note (G4), reconstructed with several methods in the oracle scenario.}
	\label{fig:ssep_piano}
\end{figure}

We illustrate these results on the same example as in the previous Section (mixture of overlapping piano notes). We observe in Fig.~\ref{fig:ssep_piano} that the \textbf{PU-Iter} estimate better fits the ground truth than the other methods. This is due to the fact that \textbf{Wiener} and \textbf{Cons-W} modify the target magnitude when sources overlap in the TF domain, which is not a desirable property if the magnitude has been reliably estimated.

Finally, it is important to note that \textbf{Cons-W} is computationally costly: for $10$ seconds excerpts, the average phase retrieval time is $11.8$ seconds with \textbf{Cons-W} vs $4.9$ seconds with our method. The proposed approach then appears appealing for an efficient audio source separation task, notably in terms of interference rejection.

\subsection{Impact of the onset phase}
\label{sec:exp_onset}

Finally, we evaluate the room for improvement of onset phase recovery. We run the \textbf{PU-Iter} procedure in the oracle scenario considering two different settings: onset phases can be estimated by assigning the mixture phase to each component (as in the previous experiment), or alternatively, they are equal to the ground truth phase to which a random error is added under the form of a centered Gaussian white noise with standard deviation $2\pi \epsilon$, where $\epsilon$ ranges between $0$ and $1$.

\begin{figure}[t]
	\hspace{-1em}
	\includegraphics[scale=0.5]{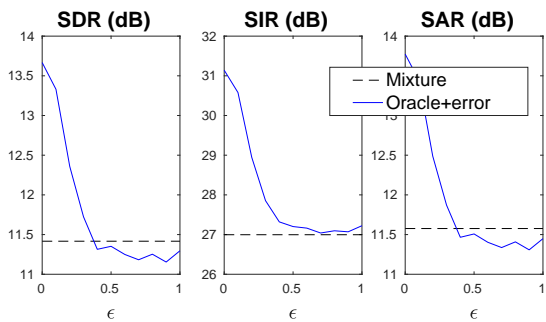}
	\caption{Separation performance for various onset phases.}
	\label{fig:compare_onset}
\end{figure}

From the results in Fig.~\ref{fig:compare_onset} we remark that there is a gap in terms of SDR, SAR ($\approx 2$ dB) and SIR ($\approx 4$ dB) between using the mixture phase and the oracle phase within onset frames. When some perturbation is added to the oracle phase, the performance decreases. We also note that in terms of SDR and SAR, when the error is of approximately 40~\%, the performance becomes similar to that of giving the mixture phase to each component within onset frames. This is consistent with the fact that the error between true phases and mixture phases within onset frames amounts to roughly $40$ $\%$. Note that this observation does not hold for the SIR. One possible explanation is that using the mixture phase yields estimates with more interferences because the mixture phase contains information that is relative to all sources. Thus, when the error between the mixture phase and the true phase is important (more than $50$ $\%$), using the mixture phase reduces the artifacts and distortion, but it may introduce some interferences.

Overall, giving the mixture phase to each component is fast and easy to implement, but onset phase reconstruction can be improved to fully exploit the potential of the PU technique.

\section{Conclusion}
\label{sec:conclu}

The source separation procedure introduced in this paper exploits the PU algorithm in order to promote a form of temporal continuity in its output estimates. The experimental results have shown that such a procedure outperforms consistent Wiener filtering in a scenario in which the magnitude spectra are known. In a more realistic scenario, where the magnitudes are estimated beforehand, it reaches a performance similar to other methods in terms of objective criteria, with a significant improvement in terms of computational cost. Sound excerpts also show that in terms of perceptual quality, this approach compares favorably with the state-of-the-art consistency-based source separation approach, which is confirmed by the computation of perception-related metrics.

The proposed approach has shown good results when the magnitude spectrograms are reliably estimated: therefore, it could be further combined with deep neural networks since these models have demonstrated remarkably good performance for magnitude estimation~\cite{Takahashi2017}. Another promising future research direction is to combine consistency-based and model-based phase recovery techniques for exploiting the full potential of both approaches, as first attempted in~\cite{Magron2017c}. Besides, as suggested by the last experiment, onset phase recovery is an interesting research direction for improved sounding quality. For instance, onsets can be modeled as impulses~\cite{Magron2015a,Sugiyama2013}, or one can use a model of repeated audio events within onset frames~\cite{Magron2015c}. Finally, some phase information can be incorporated into a probabilistic model where the phase is no longer uniform, in order to yield conservative source estimates~\cite{Magron2017}.

\bibliographystyle{IEEEtran}
\bibliography{IEEEabrv,references}

\end{document}